# HIGH-POWER TARGETS: EXPERIENCE AND R&D FOR 2 MW*

P. Hurh, FNAL, Batavia, IL 60510, USA
O. Caretta, T. Davenne, C. Densham, P. Loveridge, , STFC-RAL, Didcot, OX11 0QX, UK
N. Simos, BNL, Upton, NY 11973, US

*Abstract*

High-power particle production targets are crucial elements of future neutrino and other rare particle beams. Fermilab plans to produce a beam of neutrinos (LBNE) with a 2.3 MW proton beam (Project X). Any solid target is unlikely to survive for an extended period in such an environment - many materials would not survive a single beam pulse. We are using our experience with previous neutrino and antiproton production targets, along with a new series of R&D tests, to design a target that has adequate survivability for this beamline. The issues considered are thermal shock (stress waves), heat removal, radiation damage, radiation accelerated corrosion effects, physics/geometry optimization and residual radiation.

## INTRODUCTION

The LBNE Neutrino Beam Facility conceptual design for a future 2+ MW upgrade includes targeting 60-120 GeV pulsed proton beam from the Project X accelerator (1.6e14 protons per pulse, 1.5-3.5 mm sigma radius, 9.8 micro-sec pulse length) on a low density, solid target for the production of low energy neutrinos. Solid targets under that level of particle beam flux are extremely challenging to design, build and operate. Experience from targeting operations at FNAL's Anti-proton source and NuMI target hall have indicated the following critical design issues: thermal shock (stress waves), heat removal, radiation damage, radiation accelerated corrosion effects, physics/geometry optimization and residual radiation. Consideration of these critical design issues has resulted in a program of R&D efforts focused on two of the most promising high beam power neutrino target materials, graphite and beryllium. An overview of the critical design issues and these target material R&D activities, as well current status and preliminary results, are presented here.

## CRITICAL DESIGN ISSUES

The six critical design issues for solid, high power targets described below cannot be addressed independently. Instead, the R&D and design process must encompass all issues to arrive at a successful balance or compromise that satisfies design goals.

### Thermal Shock

Energy deposited in the target material by the high intensity primary beam over a short time scale creates a volume of heated material surrounded by cooler material. The resulting sudden compressive stress creates stress waves radiating out from the central beam spot. These stress waves reflect from free surfaces and can constructively interfere to create stress concentrations. Simulations have shown that dynamic stresses can be double that of static stresses alone depending upon the target material and characteristic length. LBNE studies predict temperature increases of over 200 K per pulse and dynamic stress beyond the yield strength (250 MPa) for a simple beryllium rod exposed to 2.3 MW of proton beam.

Methods to overcome thermal shock effects include material selection (low specific heat, low coefficient of thermal expansion, low modulus of elasticity, and high tensile/fatigue strength), segmenting target length (to avoid accumulation of expansion), avoidance of stress concentration shapes (such as sharp corners), compressive pre-loading to reduce tensile stresses, and manipulation of beam parameters (namely beam spot size and particles per pulse) to reduce stresses to tolerable levels. Designs and simulations should consider the worst case accident conditions that include maximum beam intensity, minimum spot size, and mis-steered beam.

Thermal shock is detrimental to liquids as well. Not only for liquid targets in which cavitation from the incident beam can occur, but also for cooling media in pipes positioned in the secondary shower near the target. For example, sudden temperature increases of 5C have been estimated to cause pressure rises up to 350 psi in the NuMI low energy water cooling circuit (so-called "water hammer" effect).

### Heat Removal

Energy deposited in the target material must obviously be removed to avoid unacceptably high temperatures. Typically, for neutrino targets, the fraction of beam power deposited in the target material is relatively low (25-30 kW for 2 MW primary beam). Water cooling can easily handle this level of heat removal. However, in addition to the "water hammer" problem described previously, water cooling also brings with it the problems of tritium and hydrogen gas production. Other methods of cooling such as high mass flow gaseous helium and spray cooling have advantages if acceptable heat transfer rates can be achieved.

### Radiation Damage

Although it may be fairly straightforward to design target components to stay within the known design limits of materials, it is much more difficult to confidently design for target survival in the irradiated state. As materials are irradiated their material properties change due to displacements of atoms in the crystal structure. The manner in which the damage manifests in the material properties varies depending upon the material, the initial material structure, the type of radiation, and the irradiation environment (especially irradiation temperature). Many common structural materials, such as stainless steel, can withstand 10 DPA (displacements per atom) or more before reaching end of useful life. However other materials, such as graphite, suffer significant damage at doses as low as 0.1-0.2 DPA. Many studies have been conducted over the past 60 years to determine irradiated properties for materials used in the nuclear power industry. Unfortunately for the high power target designer, such data is from neutron radiation and not high energy proton radiation. Gas production and other effects present in proton irradiation may be responsible for significant differences in radiation damage. Studies are currently underway that will hopefully shed light on this issue.

### Radiation Accelerated Corrosion

Oxidation of target materials is generally degrading to the material structure, creating initiation sites for cracks and loss of target material. For materials with high oxidation rates (such as graphite), this is overcome by operation of the target in an inert atmosphere requiring a sealed vessel with beam windows. In addition to classical oxidation, in an irradiated environment, normally stable material surface chemistry can become unstable due to the combination of radiation damage and the presence of aggressive compounds created by beam ionization of air surrounding the target. For example, aluminum materials normally have a uniform, thin oxide layer that prevents further oxidation in the presence of humid air. However, when irradiated, the formation of nitric acid and ozone from air ionization combine with radiation damage at the surface resulting in accelerated oxidation with a concerning, pit-like morphology [1]. Radiation accelerated corrosion of this type was seen on the NuMI decay pipe window and prompted a significant change in operation mode for the experiment.

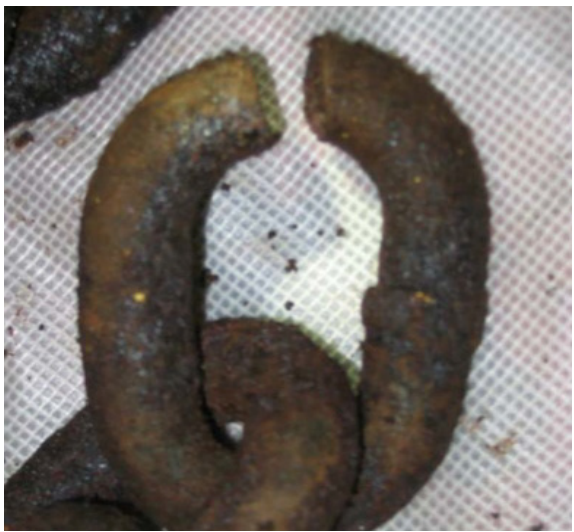

Figure 1: Broken high strength steel chain due to radiation accelerated corrosion induced hydrogen embrittlement.

Nitric acid formation in humid air exposed to beam can also accelerate corrosion of metals in the target area. With hardened steel alloys susceptible to hydrogen embrittlement this can result in premature, sudden cracking. Radiation accelerated corrosion of this type was seen in failures of high strength steel chain in the Mini-BooNE absorber (see Figure 1) and many failures of high strength steel bolts and washers in the NuMI Target Hall.

### Physics/Geometry Optimization

In neutrino targets, it is desired to maximize neutrino yield of a particular energy range per incident primary proton. However this runs somewhat counter to the thermo-mechanical requirements of a robust target design. For instance, to first order, neutrino yield increases with smaller beam spot size, yet a smaller spot size increases the thermal shock seen by the target. Similarly, the outer transverse dimension of the target should be kept small to reduce re-absorption of secondary pions, yet this reduces the structural rigidity of the target and reduces the surface area available for cooling. So, an iterative process is needed to make design changes to the target geometry/materials and then evaluate the physics and mechanical performance. To speed this iterative process, it is helpful for the designer to have a relatively simple figure of merit of physics performance to optimize rather than wait for computing intensive simulations of entire beamline optics. This proved extremely useful in LBNE beryllium target studies described later. This physics/geometry optimization can result in some rather creative and novel design ideas such as spherical targets or multi-material targets.

### Residual Radiation

Although most high power targets are designed to be replaced at their end of life rather than repaired, experience at Fermilab has shown that the ability to repair or even autopsy failed target components should be considered in the initial design. Often, the target itself does not fail, but the supporting components/systems do fail (such as a cooling circuit). Since spare targets typically are expensive and time consuming to produce, it is not uncommon for a failure to occur when there is no ready spare component. Thus repair is the only option. Although one cannot design for every eventuality, it is possible to design for easy accessibility to fasteners, clamps, and ports and utilize features easily manipulated by remotely operated tools. LBNE target component dose rates are predicted to be 100-800 R/hr on contact after 10 days of cool-down. Hands-on repair activities will be severely limited.

## LBNE GRAPHITE TARGET R&D

Graphite has been chosen as a target material for many neutrino beam facilities (NuMI, T2K, CNGS) because of its excellent resistance to thermal shock and other advantages for neutrino production. However, graphite exhibits radiation damage that changes its material properties significantly at relatively low dose.

Figure 2 shows the significant decrease in thermal conductivity of two types of graphite and a carbon-carbon composite (CX-2002U) exposed to neutron irradiation [2]. Moreover, with increased gas production associated with high energy proton irradiation (relative to neutron irradiation), the effects on graphite structure may be more severe as demonstrated by irradiation tests of graphite at BLIP (BNL) in 2006 [3]. Figure 3 shows a set of graphite samples from the 2006 BLIP test completely destroyed in the central beam spot area after an integrated flux level of ~0.5-1e21 protons/cm$^2$. This level of structural damage at relatively low dose is obviously of great concern when considering graphite as a candidate target material.

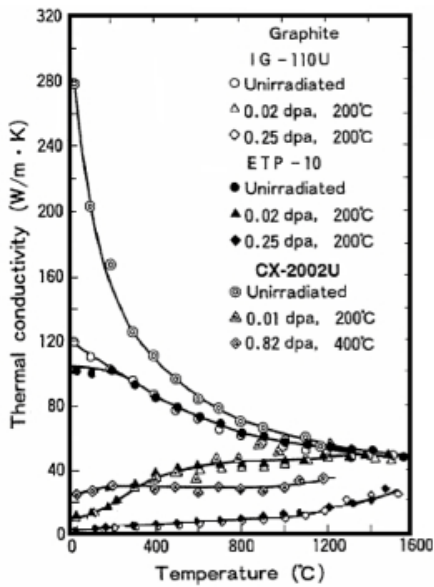

Figure 2: Effect of neutron irradiation on thermal conductivity of 3 types of carbon samples [2].

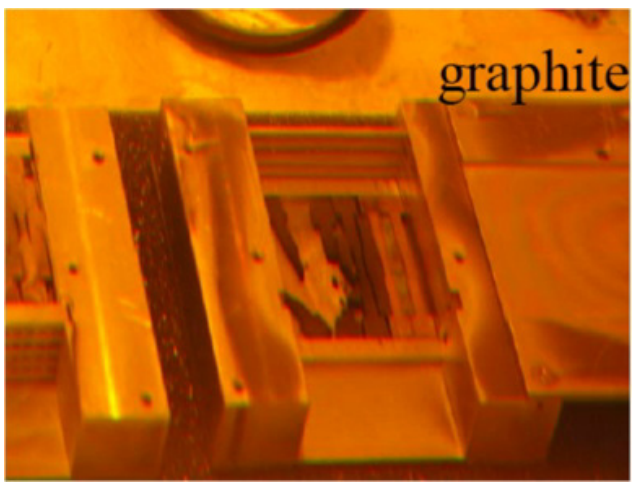

Figure 3: Graphite samples after irradiation at BLIP facility in 2006 (photo courtesy of N. Simos).

In order to further explore the structural degradation of graphite under high energy proton beam, a new test program was undertaken at the BLIP facility at BNL under the guidance of N. Simos. In this test, several grades of graphite were exposed to 181 MeV proton beam at BLIP. However, unlike the earlier BLIP tests where cooling water was in direct contact with the samples, most of the new samples were encapsulated in stainless steel containers purged with argon gas. One set of samples in this new test was installed in the water without a capsule so a direct comparison could be made between samples in a water environment and samples in an argon environment.

Table 1 lists materials tested along with the primary motivating reasons. The samples received a peak integrated flux of about 5.9e20 protons/cm$^2$ from the BLIP beam. This is about half of the integrated flux in earlier BLIP tests. Visual inspection revealed little evidence of structural degradation of any graphite samples within the argon filled capsules.

Table 1: BLIP Test Materials

| Material | Motivation |
|---|---|
| C-C Composite (3D) | 2006 BLIP failure |
| POCO ZXF-5Q | NuMI/NOvA target material |
| Toyo-Tanso IG-430 | Nuclear grade for T2K |
| Carbone-Lorraine 2020 | CNGS target material |
| SGL R7650 | NuMI/NOvA baffle material |
| St.-Gobain AX05 h-BN | Hexagonal Boron Nitride |

Figure 4 shows a post-irradiation picture of the carbon-carbon composite samples that were immersed in the water cooling medium while being irradiated. The central beam spot area was damaged with broken fibers exposed and carbon powder granules flaking off the surface. This damage on the directly water cooled samples while none was observed on the argon encapsulated samples indicates that the damage shown in the earlier BLIP tests was due, at least partially, to the water environment.

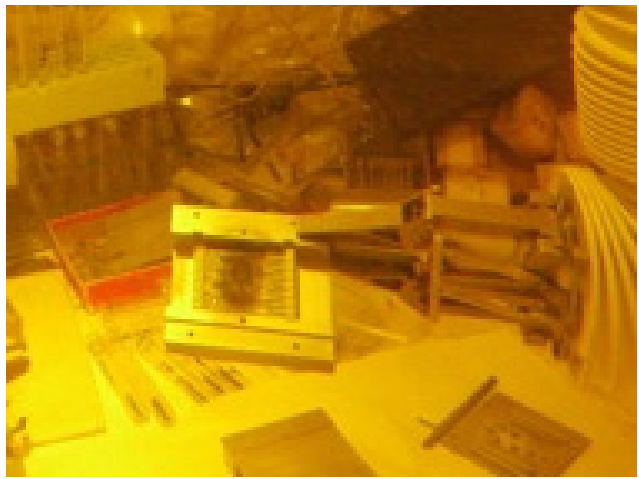

Figure 4: Water immersed C-C composite samples after irradiation at BLIP showing damage.

Figure 5 shows the thermal deflection response of a graphite sample after irradiation. Upon the first cycle to 300° C, the sample showed a significant decrease in thermal expansion. In subsequent cycles, the graphite appeared to have recovered its un-irradiated expansion characteristics. However, when compared to a control sample, the measured coefficient of thermal expansion

(CTE) is actually higher than the un-irradiated sample. This behavior was qualitatively similar across all the graphite types as shown in Figure 6.

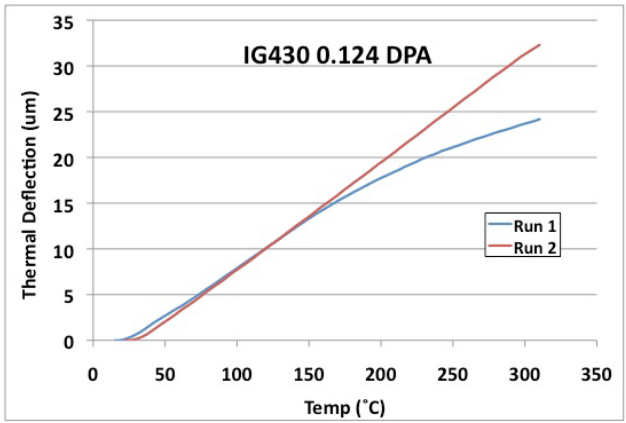

Figure 5: Expansion of IG-430 graphite during two consecutive thermal cycles after irradiating to 0.124 DPA.

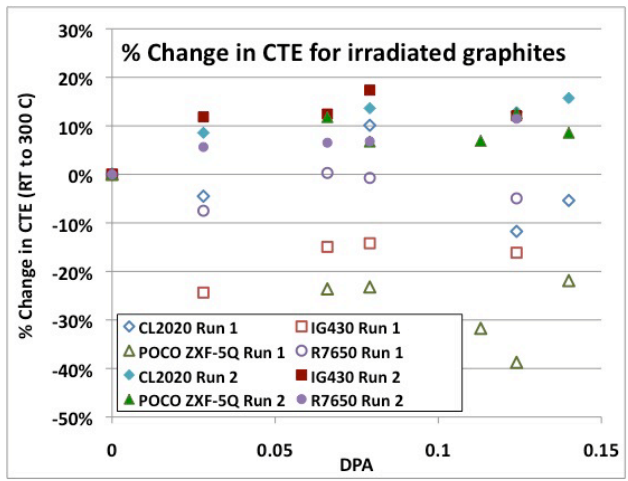

Figure 6: Comparison of change in CTE (20-300°C) for graphite samples during two consecutive thermal cycles after irradiating at BLIP (Open symbols: First cycle; Filled symbols: Second cycle).

This behavior seems to be significantly different from past studies with graphite exposed to fast neutron irradiation. In particular, the CTE under neutron irradiation was shown to increase at these low dose levels [4], contrasted with the decrease seen here. In addition, it was shown that the neutron radiation induced damage was not completely reversed unless high annealing temperatures were achieved (>1,000 C) compared to the lower annealing temperatures used in this study (300°C) [4]. Since the CTE measurements shown here seem to be consistent with the neutron irradiation damage results only after thermal cycling, perhaps there is some other damage mechanism associated with proton irradiation (such as gas production) that releases upon the first thermal cycle revealing the more permanent damage consistent with neutron irradiation. Certainly more work is needed in this area before coming to any conclusions.

Testing of the irradiated samples is continuing. Tensile testing should be underway by the time of this publication. Full results should be available by the end of 2011.

## LBNE BERYLLIUM TARGET R&D

Due to concerns over radiation damage and resulting target lifetimes, efforts to qualify beryllium as a target material were undertaken. A design study was commissioned with STFC-RAL's High Power Targets Group to explore the use of beryllium as an LBNE target for both the 700 kW and 2.3 MW primary beam powers within the parameter space listed in Table 2.

Table 2: Beam parameters for Be design study.

| Energy (GeV) | Protons per Pulse | Rep. Period (sec) | Beam Power (MW) | Beam sigma (mm) |
|---|---|---|---|---|
| 120 | 4.9e13 | 1.33 | 0.7 | 1.5-3.5 |
| 60 | 5.6e13 | 0.76 | 0.7 | 1.5-3.5 |
| 120 | 1.6e14 | 1.33 | 2.3 | 1.5-3.5 |
| 60 | 1.6e14 | 0.76 | 2 | 1.5-3.5 |

Analysis included modeling the physics in FLUKA to calculate energy deposition and simulating the thermal and structural (static and dynamic) effects in ANSYS and AUTODYN. In addition, FLUKA was used to gauge the effect of target/beam geometry variations on particle production.

Figure 7 shows a representative contour plot of equivalent stress resulting from a single pulse of 700 kW primary beam. Table 3 shows the static analysis results for various cases of beam power and target geometry. With the yield strength of Be about 270 MPa°C[5], the smaller beam spot cases (1.5 mm radius sigma) are not viable at the higher beam powers (2 and 2.3 MW). Whereas, the larger beam spot cases (3.5 mm radius sigma) are viable even at the higher beam powers.

When dynamic effects are included however, the peak stresses in the target almost double due to longitudinal stress-wave propagation. For instance, for the 2.3 MW, 120 GeV, 3.5 mm sigma case, the peak stress is 173 MPa compared to 88 MPa for static analysis alone. Since the dynamic stresses are due to longitudinal stress-waves, segmenting the target into shorter segments can reduce the resulting stresses. Figure 8 shows equivalent stress in a 50 mm long segment under the same beam conditions. It can be seen that stresses have been reduced to 109 MPa.

The effect of mis-steered beam on a beryllium target rod was simulated. Figure 9 shows that, for the 2.3 MW case, the free end of the target deflects more than 12 mm for an offset of 2 sigma. Since the LBNE target is surrounded by the focusing horn inner conductor with a clearance of 5 mm, this is clearly not acceptable. In addition, bending stresses arising from this off-center beam case exceed comfortable stress limits. Certainly adding transverse support points and segmenting the target should reduce this effect.

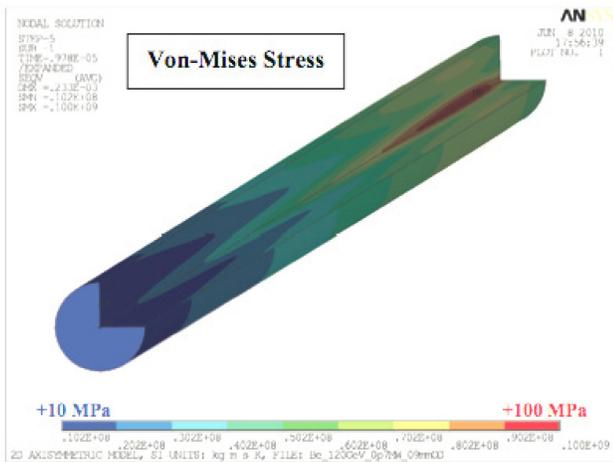

Figure 7: Equivalent stress in Be target rod from 1 pulse of 700 kW beam (static only).

Table 3: Beryllium Target Rod Static Analysis Results.

| Beam Energy & Power (GeV, MW) | Beam Sigma (mm) | Peak Energy Density (J/cc/pulse) | Max ΔT per pulse (K) | Max VM Stress (MPa) |
|---|---|---|---|---|
| 120, 0.7 | 1.5 | 254 | 76 | 100 |
| 120, 0.7 | 3.5 | 74 | 22 | 27 |
| 60, 0.7 | 1.5 | 243 | 73 | 99 |
| 60, 0.7 | 3.5 | 61 | 18 | 23 |
| 120, 2.3 | 1.5 | 846 | 254 | 334 |
| 120, 2.3 | 3.5 | 245 | 74 | 88 |
| 60, 2 | 1.5 | 707 | 212 | 288 |
| 60, 2 | 3.5 | 176 | 53 | 68 |

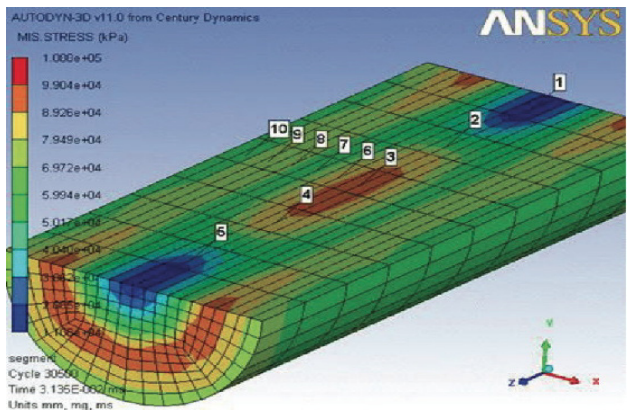

Figure 8: Equiv. stress in Be target segment from 1 pulse of 2.3 MW beam (static and dynamic).

An interesting and novel target design concept to come out of this work is pictured in Figure 10. In this case the target is segmented into spheres wrapped with helical fins to direct high mass flow gaseous helium around the spheres for cooling. Not only does the segmentation work to reduce longitudinal stresses, but the spherical shape allows pions created in the center of the target to escape without being re-absorbed into the surrounding material while also allowing coolant to flow closer to the hottest areas of the target. Although significant work was done to demonstrate the viability of this concept, much more development work is required to fully evaluate this concept.

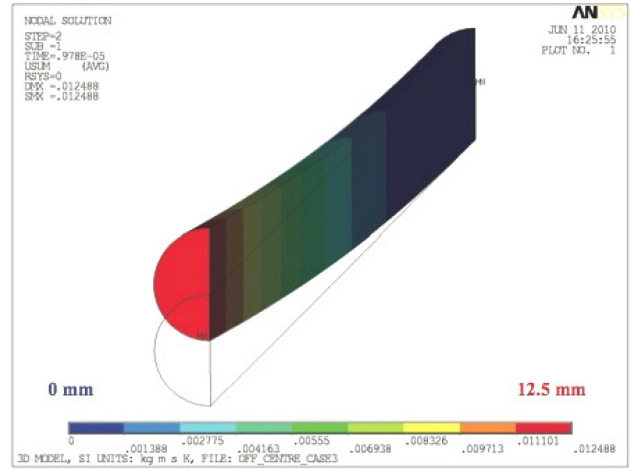

Figure 9: Deflection of Be target rod in response to a 2 sigma offset beam pulse (2.3 MW case).

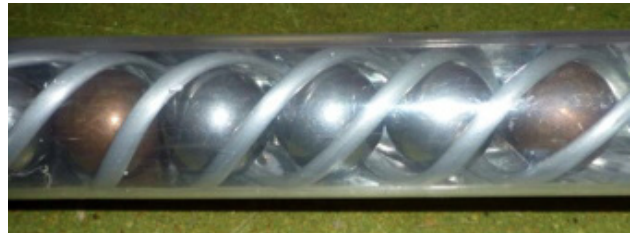

Figure 10: Mock-up of a conceptual target design using an array of spheres with helical flow guides.

## FUTURE WORK

Both graphite and beryllium remain viable as candidate high power target materials for LBNE. Near term results from the BLIP irradiation tests will shed light on the longevity of graphite in high intensity proton beam. Simulation and design work on a segmented beryllium target and cooling system should continue in the near future that includes validation of simulation methods to predict beam induced failure in beryllium